# Simple learning models can illuminate biased results from choice titration experiments

Abran Steele-Feldman

*Interdisciplinary program in Quantitative Ecology and Resource Management,
University of Washington*

and

James J. Anderson

*School of Aquatic and Fishery Sciences,
University of Washington
Box 358218, Seattle, WA 98195.
E-mail: jjand@uw.edu*

## Abstract

The choice titration procedure presents a subject with a repeated choice between a standard option that always provides the same reward and an adjusting option for which the reward schedule is adjusted based on the subject's previous choices. The procedure is designed to determine the point of indifference between the two schedules which is then used to estimate a utility equivalence point between the two options. Analyzing the titration procedure as a Markov birth death process, we show that a large class of reinforcement learning models invariably generates a titration bias, and that the bias varies non-linearly with the reward value. We treat several titration procedures, presenting analytic results for some simple learning models and simulation results for more complex models. These results suggest that results from titration experiments are likely to be biased and that inferences based on the titration experiments may need to be reconsidered.

## Introduction

When animals or humans choose between a reward delivered immediately ($5 today) and a reward delivered after some delay ($50 tomorrow), they prefer larger rewards and shorter delays. However, if the delay becomes large enough a small immediate reward will eventually be preferred to a large delayed reward. This phenomenon is known as delay discounting, and is commonly modeled by invoking a utility function that converts the amount and delay of a given reward into a univariate currency, *subjective utility*, which serves as the basis for decision making. The hyperbolic function (Mazur, 1984) is probably the most frequently used utility function for delay discounting and provides a good fit to results from a variety of experiments (reviewed in Mazur, 2001).

In order to fit and test different utility functions, many experiments attempt to find *utility equivalence points*, pairs of reward/delay combinations with the same subjective utility, which can be used to parameterize different utility functions. The titration procedure, introduced by Oldfield (1949), has been used repeatedly to estimate utility equivalence points (Acheson, Farrar, Patak, Hausknecht, Kieres, & Choi, et al. 2006; Bateson & Kacelnik, 1995, 1996; Grace, 1996; Green, Myerson, Holt, Slevin, & Estle, 2004; Lea, 1976; Mazur, 1984,1985, 1986a, 1986b, 1988, 1991, 1995, 1996, 2000, 2005; Mazur & Coe, 1987; Mazur, Synderman, & Coe, 1985; Reynolds, De Wit, & Richards, 2002; Richards, Mitchell, Dewit, & Seiden, 1997). The idea is to present an experimental subject with a choice, on discrete trials, between two options that provide different reward/delay combinations and then incrementally adjust the reward or delay provided by one of the options until the subject is equally likely to choose each option. This procedure has been used frequently by Mazur and colleagues in order to analyze delay discounting, and as a result it is strongly associated with Mazur's hyperbolic utility function (Mazur 1984; Richards et al. 1997). It has also been used to evaluate how choice depends on additional reward characteristics, such as the type of reward (Belke & Kwan, 2000).

In a traditional titration experiment, the two options are represented by different operant keys or levers: a *standard option* always provides rewards with the same amount/delay combination and an *adjusting option* provides a reward whose magnitude changes over the course of the experiment based on the subject's choices. Only one reward dimension, the *titrating dimension,* changes during an experiment. When the reward amount changes, the procedure is known as the *adjusting amount procedure* (Richards et al. 1997), and when the delay changes it is called the *adjusting delay procedure* (Mazur 1987). Eventually the magnitude of the titrating dimension comes to oscillate around a fixed value, the *indifference point*, which is used to estimate the utility equivalence point.

The titration procedure will only be a valid method for testing utility functions if the indifference points obtained with the titration procedure are unbiased estimates of the true utility equivalence points. Surprisingly, the validity of the titration procedure has received very little attention, and several of the studies that have analyzed its validity have found that it often generates biased results (Bateson & Kacelnik, 1995; Mazur, 1984). This bias has generally been attributed to an inherent bias on the part of the experimental subjects either for (Grace, 1996; Mazur 1986b) or against (Mazur, 1984) the titrating option, or as a manifestation of risk sensitivity (Bateson & Kacelnik, 1995; Mazur, 1988). However, the possibility remains that the subjects are unbiased and instead the procedure itself is responsible for the biased results. That is, the titration procedure may generate biased estimates of the utility equivalence points even if the subjects are in fact unbiased.

In this paper, we show analytically that the titration procedure will in fact generate biased results for many simple learning models even if they are unbiased. Using simulations we then demonstrate that the bias will persists for more realistic models. Since the procedure is biased for these simple models, there is little reason to expect that it will be unbiased for more complicated or realistic models of decision making. This fact suggests that inferences based on titration experiments may need to be reconsidered in light of this bias. Moreover, the magnitude and nature of the titration bias contains information about the decision processes involved, and it can thus be used to evaluate the suitability of different learning models.

The next section introduces a modeling framework, founded in reinforcement learning, which formalizes learning models using three pieces: utility functions, estimators, and choice functions. We then formally introduce one version of the titration procedure and proceed to establish conditions under which it generates a bias for some simple models. Using the results of this analysis, we numerically explore the magnitude of the bias, and use simulations to analyze more complicated learning models that are not amenable to an analytic treatment. Finally, we examine two other versions of the titration procedure, and compare the results of all three procedures.

# Reinforcement Learning Models

## Modeling Framework

The problem confronting a learning organism is often modeled as a multi-armed bandit problem (Narendra & Thathachar, 1989). The agent interacts with an external environment that presents a set of $k$ available actions $a_i$. On discrete trials $n$ the agent selects exactly one action $a(n)$ and in response receives a reward $\mathbf{x}(n)$. This framework is especially suited for modeling discrete trial operant conditioning experiments and has a long history in psychology. Titration experiments offer $k = 2$ possible actions: $a_S$ (select Standard option) and $a_J$ (select adJusting option), and each reward is characterized by an amount $x_A$ and a waiting time or delay to reward delivery $x_W$: $\mathbf{x} = \{x_A, x_W\}$.

A reinforcement learning model specifies a mechanism for selecting the next action based on the sequence of rewards obtained in previous trials. Several authors (Sutton & Barto, 1998; Yechiam & Busemeyer, 2005; Yechiam, Goodnight, Bates, Busemeyer, Dodge, Pettit & Newman, 2006) have noted that most learning models can be implemented and analyzed using a three part structure. The elements may have different names, but functionally they play similar roles. First, a utility function defines how the agent values individual rewards, mapping each multi-dimensional reward $\mathbf{x}(n)$ to the subjective utility of the reward $u(n)$. Second, utility values are stored in memory, which is most frequently modeled using *estimators* $\hat{u}_i(n)$ representing the agent's estimate of the utility expected from each action $a_i$. Third, a utility function selects the next action probabilistically based on the estimators' values (or more generally based on the values stored in memory). The interaction of these three elements determines the model's behavior.

### Utility functions

The utility function (UF) defines the subjective utility $u(n)$ of a reward $u(n) = U(\mathbf{x}(n))$. Each reward has two dimensions and the subjective utility is a non-negative scalar, i.e. $U : \mathbb{R}^2 \to \mathbb{R}_+$. The UF should be an increasing function of attractive reward dimensions, such as reward amount, but a decreasing function of unattractive dimensions, such as the reward delay.

Mazur's hyperbolic UF (Mazur 1984, 2005) will be used exclusively in the following analysis and is given by

$$U(\mathbf{x}) = U(x_A, x_W) = \frac{x_A}{1 + \alpha x_W} \qquad (1)$$

where $\alpha$ is a model parameter. Here $U(\mathbf{x})$ is a logarithmically-convex (log-convex) function of the delay but an affine function of the reward amount. In general, the convexity or concavity of a UF has dramatic impacts on choice behavior. For example, a strictly-concave UF will generate risk-averse behavior while a strictly-convex UF will generate risk-prone behavior (Kacelnik & Bateson, 1996).

### Estimators

The agent maintains estimators of the utility $\hat{u}_i(n)$ for $i = \{S, J\}$ expected from each action $a_i$ and updates these estimators after each trial. Although there are many ways to compute estimators, the exponentially weighted moving average (EWMA) is probably the most common updating algorithm used in psychology (Lea & Dow, 1984). Also known as a linear operator or leaky integrator, the exponentially weighted moving average updates the estimate recursively after each trial:

$$\hat{u}_i(n+1) = \begin{cases} \hat{u}_i(n) + m(u(n) - \hat{u}_i(n)) & \text{if } a(n) = a_i \\ \hat{u}_i(n) & \text{else} \end{cases} \qquad (2)$$

where $m \in (0,1]$ is the learning rate. Smaller values of $m$ correspond to slower learning and longer memories and as $m \to 1$ the length of the memory decreases and learning proceeds more quickly.

When $m = 1$ each estimator is equal to the utility of the last reward received from the associated action. Although this last sample estimator (LSE) is unrealistic, it is often assumed, either implicitly or explicitly (e.g. Bateson & Kacelnik, 1995), because it greatly simplifies the analysis of titration experiments.

*Choice Functions*

The agent uses a choice function (CF) $C: \mathbb{R}_+^2 \rightarrow [0,1]$ to compute the probability of selecting each option based on the estimators and then selects an action stochastically based on these probabilities. We write the CF as

$$C(\hat{u}_J(n), \hat{u}_S(n)) = P_J(n)$$
$$C(\hat{u}_S(n), \hat{u}_J(n)) = P_S(n)$$

where $P_i(n) = \Pr(a(n+1) = a_i)$. Here the CF gives the probability of choosing the action associated with its first argument.

We will say that a CF is *proper* if it meets the following two criteria. First, it must be strictly increasing in its first argument and strictly decreasing in its second argument, i.e.

$$C(y_1, z) > C(y_2, z) \quad \text{if} \quad y_1 > y_2$$
$$C(y, z_1) < C(y, z_2) \quad \text{if} \quad z_1 > z_2$$

Second it must be balanced (Falmagne, 2002):

$$C(y, z) = 1 - C(z, y). \quad (3)$$

Note that $C(y, y) = 0.5$ for a balanced CF.

Most commonly used CFs are either *difference-based* or *ratio-based,* and there is debate about which class is a better model for decision making (Corrado, Sugrue, Seung, & Newsome 2005; Fantino & Goldshmidt, 2000; Mazur, 2002; Savastano & Fantino, 1996). A difference-based CF depends only on the difference between its two argument $d(y, z) = z - y$. Thus, a difference based CF can be expressed in terms of another function $D: \mathbb{R} \rightarrow [0,1]$ such that $C(y, z) = D(d(y, z))$. The Boltzmann CF, also known as softmax (Sutton & Barto, 1998), is a common difference-based CF:

$$C(y, z) = \frac{1}{1 + e^{\frac{z-y}{\gamma}}} = D(d(y, z)) = \frac{1}{1 + e^{\frac{d(y,z)}{\gamma}}} \quad (4)$$

where $\gamma$ controls the slope of the function. Similarly, a ratio-based CF depends only on the ratio $r(y, z) = z / y$ of its arguments and can be rewritten in terms of the function $R: \mathbb{R}_+ \rightarrow [0,1]$ with $C(y, z) = R(r(y, z))$. A generalized matching CF (Baum, 1974) is ratio-based:

$$C(y, z) = \frac{1}{1 + \frac{z}{y}^{1/\gamma}} = R(r(y, z)) = \frac{1}{1 + r(y, z)^{1/\gamma}} \quad (5)$$

Again $\gamma$ controls the slope of the function and setting $\gamma = 1$ gives the matching CF.

Importantly, any ratio-based CF can be represented as the composition of a difference-based CF and a logarithmic transformation. That is, given any ratio based CF, $R$ there exists an associated *difference based analogue*, $D_R$, such that $R(r(y, z)) = D_R(d(\log(y), \log(z)))$. Thus, the Boltzmann CF is the difference based analogue for the generalized matching CF. In the following development we will use the difference-based analogue to extend results about difference-based CFs to ratio-based CFs.

## Analyzing Titration Experiments

*Mazur's Titration Procedure*

While many variations on the basic titration procedure exist, for simplicity we focus on the procedure introduced in Mazur (1984) and used, with minor variations, in many other experiments (e.g. Bateson & Kacelnik, 1995, 1996; Grace, 1996; Green et al., 2004; Mazur, 1985, 1986a, 1986b, 1988, 1991, 1995, 1996, 2000, 2005; Mazur et al., 1985; Wolff & Leander, 2002). The experimental apparatus consists of two operants, for example two keys or levers, and a food hopper (Fig. 1). Each trial begins with the illumination of one or both of the keys, and if the subject presses an illuminated key, it receives the associated reward after a short delay.

An experiment is organized into blocks containing $T$ trials each. The first $T - 2$ trials in a block are *forced trials* in which only one of the keys is illuminated. These are followed by two *choice trials* in which both keys are illuminated but the subject can choose only one of them; forced trials ensure that the subject is exposed to both options equally in each block before the choice trials interrogate the subject's preferences. The option presented on each forced trial is chosen quasi-randomly so that each option is presented on exactly half of the forced trials in a block (Fig. 1).

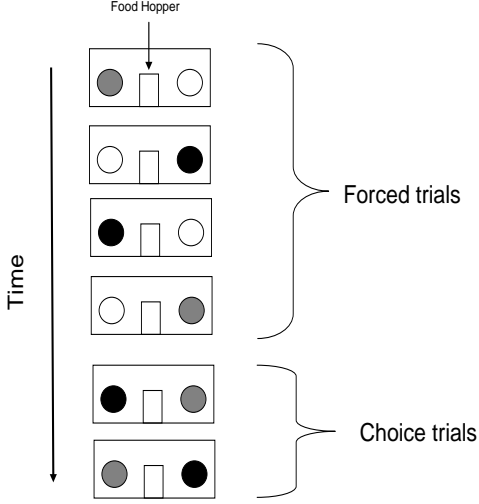

**Fig. 1**. Example of an experimental block with $T = 6$ trials. The experimental apparatus consists of two keys that can illuminate with colored lights (here shown as grey and black) and a food hopper. The first four forced trials illuminate only one of the two keys. The last two choice trials illuminate both keys. In either trial type, pressing an illuminated key delivers food to the hopper after a delay.

Throughout the experiment, the standard option always delivers the same reward on each trial $\mathbf{x}(n) = \mathbf{s}$. In a given block $N$ the adjusting option delivers the same reward on each trial $\mathbf{x}(n) = \mathbf{\tau}(N)$ but this reward can change between blocks according to the titration rule. However, only the magnitude of the titrating dimension changes during the experiment; let $x(n)$ denote the value of the titrating dimension for the reward received on trial $n$ and similarly let $\tau(N)$ and $s$ denote the value of the titrating dimension for the adjusting and standard options in each block. Assume that $\tau(N), s \in [0,\infty)$. For simplicity, we write the utility function as a function only of the titrating dimension $U(\mathbf{x}) = U(x)$ suppressing its dependence on the other reward dimension.

Either reward amount (adjusting amount procedure) or reward delay (adjusting delay procedure) can be titrated, and the titration rule must ensure that the utility of the adjusting option changes appropriately. Under the adjusting amount (delay) procedure, the titration rule decreases (increases) the value of the titrating dimension if the adjusting option is chosen on both choice trials in a block. Conversely the titrating dimension is increased (decreased) if the standard option is chosen on both choice trials. If each option is chosen once, the titrating dimension is not changed. An procedure must specify two adjusting functions $f^+, f^-$ that determine how the value of the titrating dimension is increased and decreased. Most commonly, adjustments are either arithmetic or geometric. With arithmetic adjustments

$$f^{\pm}(\tau(N)) = \tau(N) \pm \delta \qquad (6)$$

where $\delta > 0$ is the step size. With geometric adjustments the step size depends on the current value of the adjusting option. The bulk of the rest of the paper will focus on such arithmetic adjustments, but we will return address geometric adjustments towards the end of the paper.

Define the *titration probabilities* as the probability of increasing or decreasing the value of the adjusting option after block $N$. In general, the functional form of $\theta^+$ and $\theta^-$ depends on the probability of the subject choosing each of the options given the current and previous values of the titrating dimension.

$$\theta^+(N) = \Pr(\tau(N+1) = f^+(\tau(N)))$$
$$\theta^-(N) = \Pr(\tau(N+1) = f^-(\tau(N)))$$

### Titration bias

The goal of a titration experiment is to find a utility equivalence point: the titrating dimension value $\tilde{\tau}$ such that

$$U(\tilde{\tau}) = U(s). \qquad (7)$$

However, because the value at the utility equivalence point cannot be measured directly, the titration procedure is instead used to compute an indifference point, $\bar{\tau}$, which then serves as an estimate of $\tilde{\tau}$. The indifference point is computed as the average value taken by the titrating dimension towards the end of the experiment, usually after some stability criterion has been satisfied (see Mazur, 1984). For example, if the entire experiment lasted 1000 blocks, we might compute the indifference point by averaging the titrating values from the last 200 blocks. Assuming that the experiment lasts long enough and that sufficient blocks are included in the average, the expected value of the indifference point equals the asymptotic expected value of the titrating dimension $E(\bar{\tau}) = \lim_{N \to \infty} E(\tau(N))$.

If the titration procedure is a valid method for obtaining utility equivalence points then $E(\bar{\tau}) = \tilde{\tau}$. Define *titration bias* as $\Delta = E(\bar{\tau}) - \tilde{\tau}$. When might we expect a non-zero titration bias? We can think of the value of the titrating dimension as a one-dimensional random walk with time varying probabilities $\theta^-(N)$ and $\theta^+(N)$ representing the probability of taking a step to the left or right respectively; possible examples for the titration probabilities are shown in Fig. 2. There are two possible ways that a titration bias can emerge. The first is depicted by the solid line in Fig. 2; note that the probability of taking a step back towards the center increases more steeply to the left of the indifference point than it does to the right. As a result the random walker will be driven back to the center more quickly on the left than it will on the right, and it will thus tend to wander further to the right of the indifference point (Bateson & Kacelnik, 1995). Therefore, the expected position of the random walker will be to the right of the indifference point, and we would expect a positive titration bias, $\Delta > 0$. We will refer to this as a *robust* titration bias. The dashed line in Fig. 2 is symmetric around the indifference point, and thus will not generate a robust titration bias.

The second way a titration bias can emerge is due to the asymmetrical range available to the random walker: the range for the random walk is bounded below by zero but unbounded above. This asymmetry means that the system can randomly wander further to the right than it can to the left, even if the titration probabilities are completely symmetric (dashed line). Thus just by sheer chance the random walker is more likely to be to the right of the indifference point, again leading to a positive titration bias. We will refer to this as the *residual* titration bias. In the next section we derive conditions under which the titration procedure with arithmetic adjustments will generate either a robust or residual titration bias.

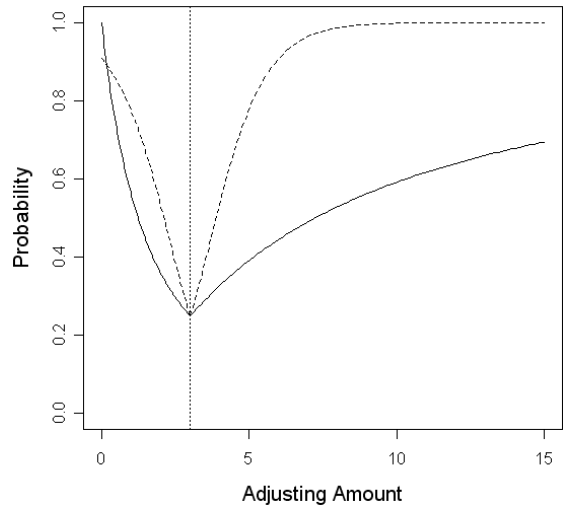

**Fig. 2.** Probability of taking a step towards the indifference point in an adjusting amount experiment as in Fig. 1. The standard provides a 3 unit reward, which represents the utility equivalence point (dotted line). The solid line was generated with the matching CF. The dashed line was generated with the exponential CF. In both cases, lines to the left of the indifference point depict $\theta^+(N)$ and to the right depict $\theta^-(N)$.

*Analytic Results*

In order to compute the expected titration bias we need to determine the asymptotic expected value of $\tau(N)$. For general learning models, this can be quite difficult. However, if the subject uses the LSE, the estimator for each option depends only on the last sample obtained from that option. Since the subject always receives at least one sample (during the forced trials) from each option in a block, it follows that the estimators on the choice trials are equal to the utility of the rewards in the current block. Thus, the adjusting option estimator is $\hat{u}_J(n) = \hat{u}_J(N) = U(\tau(N))$ and the standard option estimator is $\hat{u}_S(n) = \hat{u}_S = U(s)$.

Since the estimators only depend on the reward values in the current block, the choice probabilities, and thus the titration probabilities, also depend only on the reward values in the block. Moreover, the value of the titrating dimension is confined to a discrete set of states $i \in \{0, 1, ...\}$ where $\tau_i = i\delta$ is the value of the titrating dimension in state $i$. So we can define the titration probabilities conditional on the current state as

$$\theta_i^+(N) = \Pr\left(\tau(N+1) = \tau_{i+1} \mid \tau(N) = \tau_i\right)$$
$$\theta_i^-(N) = \Pr\left(\tau(N+1) = \tau_{i-1} \mid \tau(N) = \tau_i\right) \quad (8)$$

In the adjusting amount procedure (attractive titrating dimension), the titration probabilities are[1]

$$\theta^-(N) = \left(C(\hat{u}_J(N), \hat{u}_S)\right)^2$$
$$\theta^+(N) = \left(C(\hat{u}_S, \hat{u}_J(N))\right)^2 \quad (9)$$

and thus

$$\theta_i^-(N) = \left(C(U(\tau_i), U(s))\right)^2$$
$$\theta_i^+(N) = \left(C(U(s), U(\tau_i))\right)^2 \quad (10)$$

Similarly, with the adjusting delay procedure (unattractive titrating dimension) we have:

$$\theta_i^-(N) = \left(C(U(s), U(\tau_i))\right)^2$$
$$\theta_i^+(N) = \left(C(U(\tau_i), U(s))\right)^2 \quad (11)$$

In either case, the dynamics of the titrating dimension obey the Markov property: the distribution of $\tau(N+1)$ depends only on the value of $\tau(N)$. Thus, $\{\tau(1), \tau(2), ...\}$ defines a discrete time Markov chain with transition probabilities $P_{ij} = \Pr\left(\tau(N+1) = \tau_j \mid \tau(N) = \tau_i\right)$ given by

$$P_{ij} = \begin{cases} \theta_i^- & \text{if } j = i-1 \\ 1 - \theta_i^- - \theta_i^+ & \text{if } j = i \\ \theta_i^+ & \text{if } j = i+1 \\ 0 & \text{else} \end{cases} \quad (12)$$

with $\theta_0^- = 0$. This is known as a Markov birth-death process and asymptotically the expected value of the adjusting option (Gallager, 1996; Howard, 1971) is[2]

$$E(\bar{\tau}) = \lim_{N \to \infty} E(\tau(N)) = \sum_{i=0}^{\infty} \tau_i Y_i \Bigg/ \sum_{i=0}^{\infty} Y_i \quad (13)$$

where $Y_0 = 1$ and $Y_i = \prod_{j=0}^{i-1} \frac{\theta_j^+}{\theta_{j+1}^-}$.

In the Appendix, we use Eq. (13) to prove the following theorems:

<u>Theorem A.1</u>: With an attractive titrating dimension, models with a proper difference-based (ratio-based) CF and a strictly concave (log-concave) UF will generate a robust titration bias.

<u>Theorem A.2</u>: With an unattractive titrating dimension, models with a proper difference-based (ratio-based) CF and a strictly convex (log-convex) UF will generate a robust titration bias.

<u>Theorem A.3</u>: With either attractive or unattractive titrating dimensions, a proper difference-based (ratio-based) CF and an affine (log-affine) UF will not generate a robust titration bias.

Although these results apply to all UFs, we are most interested in their consequences for Mazur's hyperbolic UF function. As noted earlier, the hyperbolic UF is an affine function of reward amount and a log-convex function of reward delay. The preceding theorems thus imply that learning models using the hyperbolic UF, LSE, and a ratio based CF always generate a robust bias under both the adjusting amount and delay procedures. Similarly, learning models using the hyperbolic UF, LSE, and a difference-based CF generate a robust bias under the adjusting delay procedure, but only a residual bias under the adjusting amount procedure.

---

[1] The adjusting reward only changes when the subject chooses the same option on both of the choice trials in a block. Thus, when the adjusting reward is equal to, the probability of choosing the adjusting option on both choice trials in a block is equal to the square of the CF.

[2] This is not strictly true, as it is possible that $\theta_{j+1}^- = 0$ for some $j$ in which case Eq. (13) involves division by zero. In this case, the product should run from $j = \min(j : \theta_{j+1}^- > 0)$ and the summations adjusted accordingly. This minor technical point does not affect the following analysis.

*Longer Memories*

Derivation of the preceding analytic results depended heavily on the assumption that the subject uses the LSE. This assumption is unrealistic: it is likely that organisms will base their decisions on more than just the last sample. Indeed, several studies suggest that data is best fit using a learning rate parameter $m < 1$ (Cardinal, R., Daw, N., Robbins, T., & Everitt, B. (2002); Corrado et al., 2005; Lau & Glimcher, 2005). Unfortunately we cannot present analytic results for such models, but we can evaluate the resulting titration bias using simulations. Fig. 3 presents simulation results for the same experiment depicted in Fig. 1. Happily, the titration bias is bounded below by the bias computed using the LSE, and thus the analytic results serves as a lower bound to the bias for a model with a more realistic EWMA.

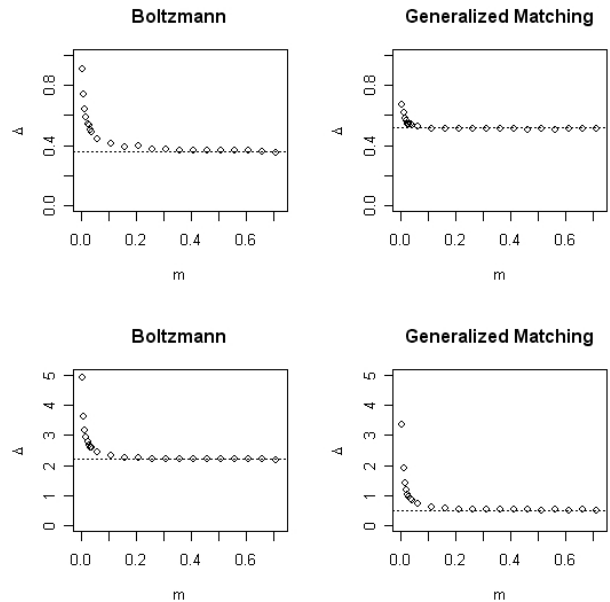

**Fig. 3.** Titration bias with longer memories. The top row shows the bias from the adjusting amount procedure where the adjusting option has a delay of 10 units and the standard amount is 5 units. The bottom row shows an adjusting delay procedure where the adjusting option delivered an amount of 5 units and the standard delay is 10 units. Each data point was computed as the mean of 100 replicates of a simulated titration experiment consisting of 300,000 blocks with 2 forced trials per block. The dashed line shows the bias for the LSE (i.e. $m = 1$) computed using Eq. (13). In all graphs $\alpha = 1$ and $\gamma = 2$.

The titration bias does not deviate substantially from the bias predicted for the LSE until the learning rate falls well below 1, but as the learning rate approaches 0, the bias increases under both procedures and CFs (Fig. 3). There are two reasons why a smaller learning rate will impact the titration bias. Firstly, decreasing the learning rate increases the length of the memory and causes the subject's estimate to lag behind the current value of the adjusting option, meaning that the subject will be slower to respond to changes in the titrating dimension. This allows the value of the titrating dimension to wander away from the indifference point and increase the magnitude and influence of the residual titration bias. Secondly, if $m < 1$ the estimate will be an average of several different values. If the UF is nonlinear this average will not be equal to the utility at the indifference, even if the titrating value oscillates symmetrically around the indifference point. Risk sensitive behavior will emerge for models with nonlinear functions due to Jensen's inequality: convex functions will generate risk prone behavior and concave functions generate risk-averse behavior (Kacelnik & Bateson, 1996). Thus, models with a convex UF will generate a larger bias relative to a model with the LSE, whereas those with a concave UF will produce a smaller bias. In the latter case, decreasing the learning rate has two opposing effects: it increases the magnitude of the residual bias but decreases the magnitude of the robust bias. The next section will explore the magnitude of the bias numerically, and we will see how these two mechanisms interact to increase the bias when the learning rate decreases.

*Magnitude of bias: Arithmetic adjustments*

Fig. 4 shows how the magnitude of the titration bias changes as a function of the UF and CF parameters, given *identical rewards* where the value of the non-titrating dimension is the same for both options. The bias magnitude increases with $\gamma$ for both CFs under both procedures. Increasing $\alpha$ also increases the titration bias with the Boltzmann CF (the value of $\alpha$ has only minimal effect on the bias with the generalized matching CF because it is effectively canceled by taking the ratio). The Boltzmann CF generates a relatively large bias with an adjusting delay but needs large values of $\alpha$ and/or $\gamma$ to generate a substantial bias with an adjusting amount.

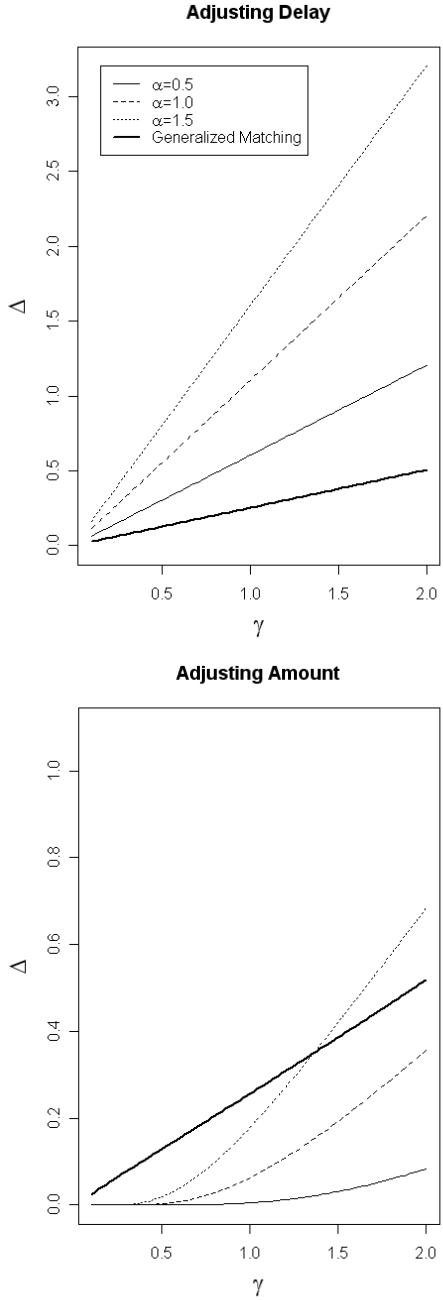

**Fig. 4.** Titration bias vs. the CF parameter $\gamma$ under arithmetic adjustments. Thin lines show Boltzmann CF and thick line shows generalized matching CF. In both, cases, the standard option provides a 5-unit reward amount after a 10-unit delay. The adjusting option provides a 5-unit reward amount and an adjusting delay (top frame), an adjusting option providing a 10-unit delay and an adjusting amount (bottom frame). Only one curve is shown for the generalized matching CF because $\alpha$ had a negligible impact on the titration bias.

Fig. 5 shows how the indifference point changes with the standard's value given identical rewards. With the Boltzmann CF and the LSE, the bias declines as the standard increases under the adjusting amount procedure, but under the adjusting delay procedure the bias increases approximately linearly as the standard increases. In the former case, the bias is completely residual, and as the value of the standard increases the residual bias goes to zero, leading to an almost unbiased indifference points. With the generalized matching CF and LSE the bias is small and virtually constant under both procedures, although it decreases slightly as the standard increases.

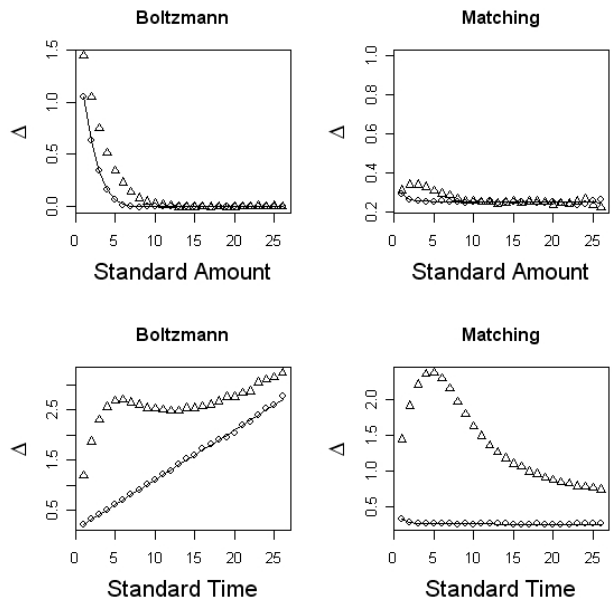

**Fig. 5.** Titration bias with arithmetic adjustments and identical rewards. The x-axis depicts titrating dimension value from the standard option. Amount titrated with a 10-unit delay (top row) and delay is titrated with a 5-unit reward (bottom row) computations use hyperbolic UF with $\alpha = 1$. The solid line depicts bias computed with Eq. (13) for a model using the LSE, and the circles show the simulated results. The triangles show the simulated results for a learning model with $m = 0.01$. The simulated indifference points were computed as the mean over 100 simulations of an experiment consisting of 200,000 blocks with 2 forced and 2 choice trials per block. The titrating value for adjusting option was initialized to the corresponding value for the standard.

The situation is substantially different in the simulations with $m < 1$ (Fig. 5). In the adjusting amount simulations, the slower learning rate increases the magnitude of the residual titration bias, but the impact declines as the value of the standard increases. Since here the hyperbolic UF is affine with respect to the titrating dimension, Jensen's inequality and risk sensitivity do not come into play. However, with an adjusting delay the hyperbolic UF is a strictly convex function of the titrating dimension, and thus generates risk-prone behavior which increases the bias with respect to the LSE. Thus a slower learning rate increases both the robust and the residual titration bias under adjusting delays, but only increases the residual titration bias under adjusting amounts.

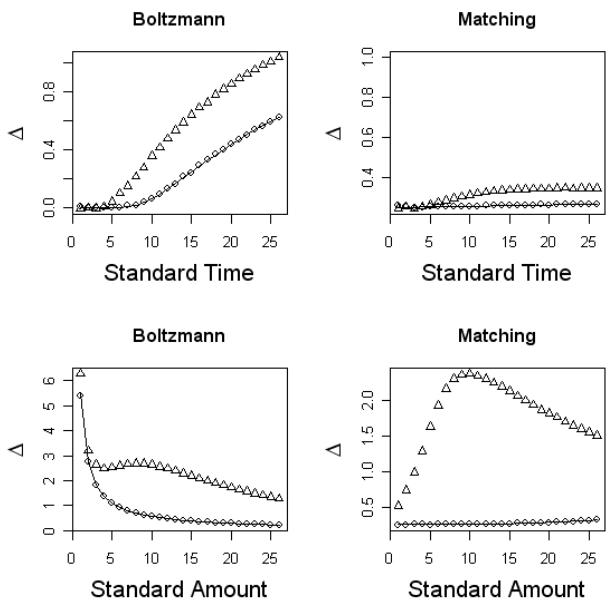

**Fig. 6.** Titration bias with arithmetic adjustments and non-identical rewards. The x-axis shows the value of the non-titrating dimension from the standard option. The top row shows the bias from an adjusting amount procedure where the adjusting option has a delay of 10 and the standard amount is 5. The bottom row shows an adjusting delay procedure where the adjusting option delivered an amount of 5 and the standard delay is 10. In all graphs $\alpha = 1$. All symbols and simulation details are as in Fig. 5.

When rewards are not identical, the titration bias cannot be evaluated empirically because the UF, and thus the equivalence point, is unknown. However, if we assume a specific UF, we can compute the utility equivalence point with Eq. (7) and then compute the expected titration bias by computing $E(\bar{\tau})$ with Eq. (13). Fig. 6 shows how the bias changes for the hyperbolic UF when the rewards are not identical. With the LSE and Boltzmann CF, the bias varies non-linearly with the value of the other standard dimension; with the LSE and the generalized matching CF the bias is approximately constant across other standard dimension values, and the magnitude of the bias is very close to that observed with identical rewards under the corresponding procedures.

With a smaller learning rate and adjusting amounts, the bias is qualitatively similar to that observed with the LSE, albeit a bit larger due to an increase in the residual bias. Under adjusting delays, however, the bias change non-linearly, just as it did under identical rewards, due to the risk-preference generated by the convex UF.

*Comparison with empirical results*

Fig. 7 shows the bias from a set of experiments using arithmetically adjusting amounts. Note that these experiments utilize different types of subjects and slightly different procedures. Most experiments show a positive bias, in accord with the predictions from the previous section. Mazur (1984) used several different standard values, and found that the magnitude of the bias increased approximately linearly with the value of the standard, with $\Delta/s \approx 0.15$, a result consistent with model predictions given the Boltzmann CF, but inconsistent with those observed for the generalized matching CF, at least for the parameter values depicted in Fig. 5. However, this pattern is not born out in the cross experiment analysis presented in Fig. 7, where there is an overall downward trend in the magnitude of the bias. This trend is driven mainly by three experiments (Mazur, 1986a, 1995; Grace, 1996) that produced a negative bias which is not predicted by any of the models discussed in the previous sections. While the negative bias observed in Mazur (1995) could be due to random variation, this is unlikely in the other two experiments as the bias in both Mazur (1986a) and Grace (1996) was observed across multiple standard values and for nearly every subject in the experiment. Interestingly, Mazur (1984), Mazur (1986a) and Grace (1996) all used virtually identical procedures, but the results from the latter two experiments are completely at odds with the results from the first one. These contradictory results deserve closer study, but the negative titration bias is difficult to explain given the simulation results presented in the previous section.

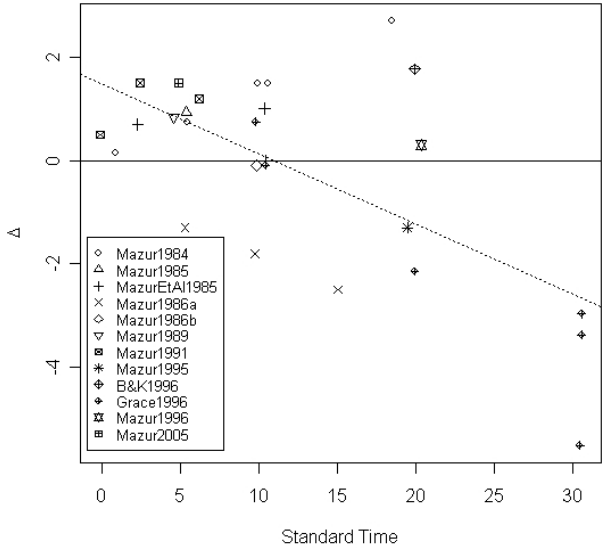

**Fig. 7**. Observed bias in experiments using arithmetically adjusting delays and identical rewards. Most experiments used pigeons, but one used starlings and one used rats. The value of the reward amount denotes either a number of food pellets or a number of seconds of access to food. All experiments used a block structure consisting of a fixed number of forced trials followed by a fixed number of choice trials. The solid line represents no bias, and the dotted line is the best fitting linear regression line. Indifference points were estimated from graphs if not reported numerically.

Only three experiments (Bateson & Kacelnik, 1995, 1996; Mazur 2000) have used an arithmetic adjusting amount procedure with identical rewards, but all three demonstrated a positive titration bias consistent with all models in the previous section. However, only one experiment utilized multiple standard values (Bateson & Kacelnik, 1995) and it showed a bias that increased approximately linearly with the standard value, $\Delta/s \approx 0.33$. This result is inconsistent with all of the models discussed in the previous section which predict either a decreasing or approximately constant relationship between the standard size and the magnitude of the bias. This incongruity suggests that a different UF or CF will be needed to explain these results. For example, a UF that is strictly concave with respect to reward amount can generate a titration bias that increases with the standard size (Steele-Feldman, 2006), so Bateson & Kacelnik's results could be taken as tentative evidence against a UF that is affine with respect to reward amount.

### Geometric Adjustments

Under arithmetic adjustments, the step size is the same in each block independent of the current value of the adjusting option, whereas under geometric adjustments the step size depends on the current value of the adjusting option. At least two different types of geometric adjustments have been used, what we will call symmetric and asymmetric. Under symmetric geometric adjustments (Lea, 1976) the adjusting functions are

$$f^{-}(\tau(N)) = \tau(N)\delta$$
$$f^{+}(\tau(N)) = \tau(N)/\delta \quad (14)$$

where $\delta \in (0,1)$. Under asymmetric geometric adjustments (Richards et al., 1997), they are

$$f^{-}(\tau(N)) = \tau(N)(1+\delta)$$
$$f^{+}(\tau(N)) = \tau(N)(1-\delta). \quad (15)$$

Note, $f^{+}(f^{-}(x)) = x$ for all $x$ under symmetric adjustments, but not asymmetric adjustments. As a result, experiments with asymmetric geometric adjustments are not confined to a discrete state space because the potentially available state space changes after each block (Fig. 12). Without a discrete state space, the analytic methods used with arithmetic adjustments cannot be applied to experiments using asymmetric geometric adjustments. However, those techniques are applicable to experiments using symmetric geometric adjustments, and in the Appendix we derive sufficient conditions for a model to generate a positive titration bias under symmetric geometric adjustments. Because the conditions are not especially illuminating, we instead focus here on how the magnitude of the bias changes given each adjusting procedure, but urge the interested reader is to check the Appendix for details.

*Magnitude of the bias: Symmetric geometric adjustments*

To our knowledge symmetric adjustments have only been used in one adjusting delay experiment (Lea, 1976) that used a relatively unorthodox procedure without forced trials wherein the adjusting value was changed after every single choice trial. Rather than explore the consequences of these procedural variations, we here examine how symmetric geometric adjustments will affect the titration bias observed within a more conventional Mazur-type block structure (2 forced trials followed by 2 choice trials).

With identical rewards (Fig. 8) and adjusting amounts, the bias is relatively small but increases approximately linearly with the value of the standard, and the learning rate has little impact. However, with identical rewards and adjusting delays the bias is overall much larger, and increases even more when the learning rate decreases. Due to the large computed biases, the simulations were run with a maximum possible titrating value of 1000, while the analytic computations were unbounded. This upper bound is responsible for the divergence between the computed and simulated results with the Boltzmann CF under adjusting delays.

With non-identical rewards (Fig. 9) the bias is again small under adjusting amounts, and the learning rate has little impact on its magnitude. Under adjusting delays the bias is much larger and increases substantially when the learning rate decreases. As with arithmetic adjustments, the relationship between bias magnitude and standard size is again quite different from that observed with identical rewards.

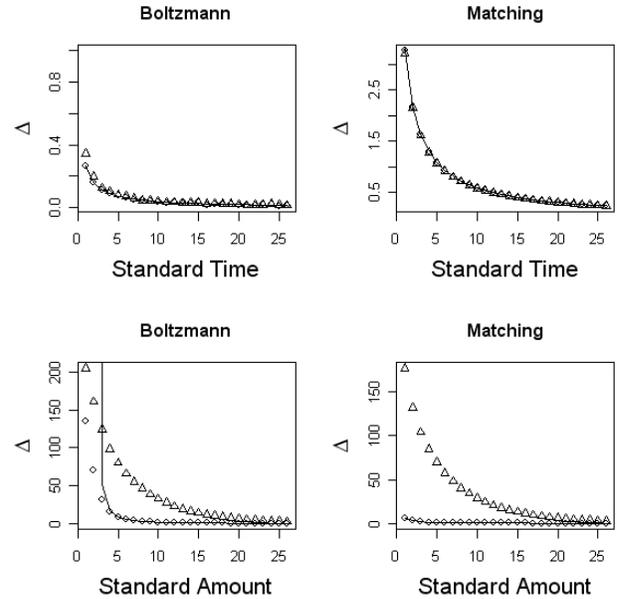

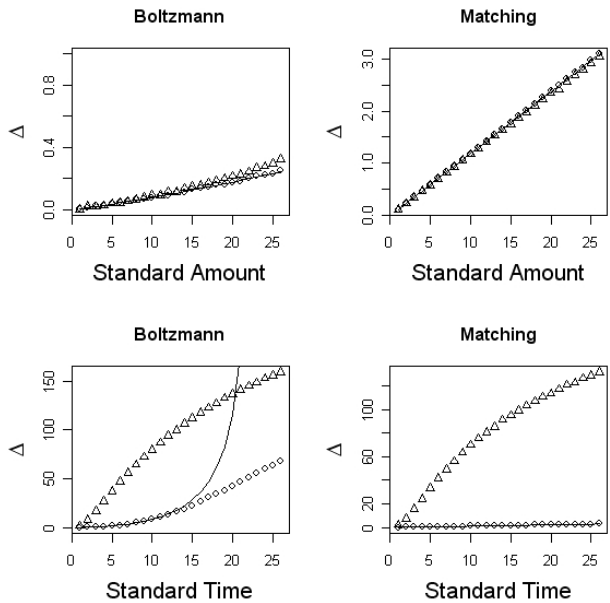

**Fig. 8**. Titration bias with symmetric geometric adjustments and identical rewards. The x-axis shows the value of the titrating dimension from the standard option. In the top row the amount is titrated with a 10-unit delay and in the bottom row delay is titrated with a 5-unit reward. All computations use the hyperbolic UF with $\alpha = 1$. The solid line depicts the bias computed with Eq. (33) for a model using the LSE, and the circles show the simulated results. The triangles show the simulated results for a learning model with $m = 0.01$. Simulation details are as in Fig. 5 except that the value of the titrating dimension was bounded above at 1000 units. If it went larger than this value, it was reset to 1000.

**Fig. 9.** Titration bias with symmetric geometric adjustments and non-identical rewards. The x-axis shows the value of the non-titrating dimension from the standard option. The top row shows the bias from an adjusting amount procedure where the adjusting option has a delay of 10 and the standard amount is 5. The bottom row shows an adjusting delay procedure where the adjusting option delivered an amount of 5 and the standard delay is 10. Details as in Fig. 8.

The one experiment to use symmetric geometric adjustments (Lea, 1976) included several conditions with identical rewards: in an exploratory phase, his pigeons showed a substantial positive titration bias, but in a following test phase that used an identical procedure his subjects showed no bias at all. Our simulation results are consistent with the positive bias observed in the exploratory phase, but unable to explain the lack of a bias in the test phase. A couple caveats are in order, however. First of all, as discussed above, the block structure used in the experiments was different than the one we simulated. Secondly, Lea (1976) reports the median value taken by the adjusting delay, but only on trials where the adjusting option was chosen, whereas we compute the mean value of the adjusting delay across all trials towards the end of the experiment. Lea's measure will be biased toward smaller indifference points if the distribution of adjusting delays is right-skewed and/or if the subjects are less likely to choose the adjusting option when the adjusting delay is large. The second of these conditions is clearly true, and the first is true for all of the simulations we have conducted (results not shown). Thus although some authors (Bateson & Kacelnik, 1996; Steele-Feldman, 2006) have suggested that symmetric geometric adjustments are perhaps more likely to produce unbiased indifference points, the present analysis suggests that the lack of a bias observed in the test phase of Lea (1976) is more likely an artifact of the unusual response measure than directly due to the adjusting procedure used.

## *Magnitude of the bias: Asymmetric geometric adjustments*

The asymmetric geometric adjustment procedure was introduced by Richards et al. (1997) in an adjusting amount experiment, and it is known as the adjusting amount procedure within the psychopharmacological literature, and has since been used in many other experiments (Acheson et al., 2006; Mitchell & Rosenthal, 2003; Reynolds et al., 2002; Richards et al. 1997; Wade, De Wit, & Richards, 2000). This procedure also utilizes a unique block structure whereby the titrating dimension is changed after every choice trial and forced trials are only presented after two consecutive choices for the same option, in which case the other option is presented in a forced trial. Because this block structure is quite common in experiments that use asymmetric adjustments, the simulation results presented below utilize it rather than the Mazur type block procedure used in the other simulations. Note, however, that some experiments have combined a Mazur-type block procedure with asymmetric geometric adjustments (e.g. Ho, Wogar, Bradshaw, & Szabadi, 1997).

With identical rewards (Fig. 10) and the LSE a negative titration bias is observed for all simulations except for the adjusting delay procedure with the Boltzmann CF where the bias is slightly positive. As with the other simulations, decreasing the learning rate increases the absolute magnitude of the bias, but in these simulations the direction of the effect depends on the dimension being titrated: under adjusting amounts the smaller learning rate makes the bias more negative, but under adjusting delays the smaller learning rate makes the bias more positive, effectively reversing the direction of the bias for the matching choice function under adjusting delays.

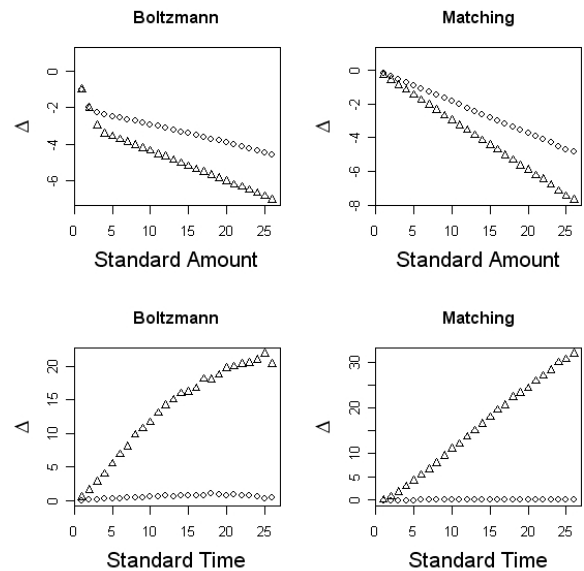

**Fig. 10.** Titration bias with asymmetric geometric adjustments and identical rewards. The x-axis shows the value of the titrating dimension from the standard option. Experimental and simulation details are as in Fig. 8. Circles show simulated results for a model using the LSE, and the triangles show the simulated results for $m = 0.01$.

With non-identical rewards (Fig. 11) the pattern is quite similar, with smaller learning rates increasing the absolute magnitude of the titration bias. The relationship between bias magnitude and standard size is again quite different from that observed with identical rewards, suggesting that Mazur's correction method is also insufficient for this experimental procedure. Note that in the simulations for the Boltzmann CF under adjusting delays, the decrease in bias at small standard values is an artifact of the upper bound at 1000 which effectively prevents the indifference point from growing as large as it would in the absence of the bound, leading to a smaller, even negative, bias.

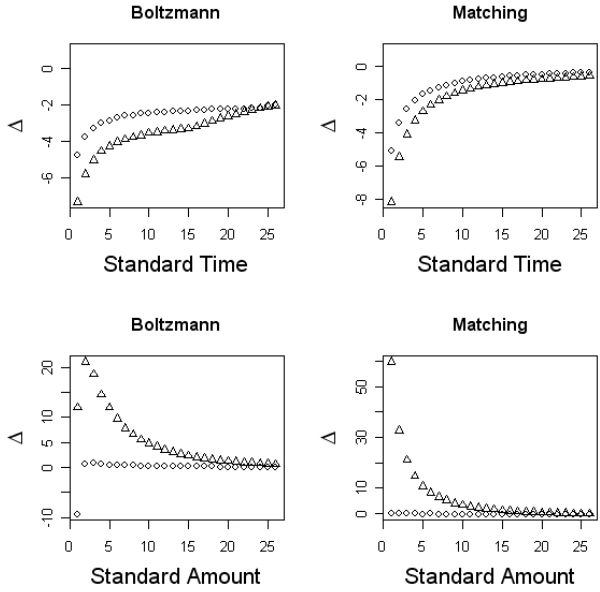

**Fig. 11**. Titration bias with asymmetric geometric adjustments and non-identical rewards. The x-axis shows the value of the non-titrating dimension from the standard option. Experimental and simulation details are as in Fig. 8. All symbols are as in Fig. 10.

The negative titration bias observed consistently under asymmetric geometrically adjusting amounts is novel, since all other procedures generated a positive titration bias. One source of this bias is depicted in Fig. 12 which shows how the states develop asymmetrically: there are more 'small' states than 'large' states. This asymmetric distribution of states will produce a negative titration bias in a manner analogous to the residual titration bias with arithmetic adjustments, namely the system has more accessible states below the indifference point than above and will thus spend more time below the indifference point.

Empirical studies using the adjusting amount procedure with asymmetric geometric adjustments and identical rewards have consistently produced a negative titration bias (Fig. 13), and the magnitude of the bias increases with the size of the standard. These empirical results are consistent with both the Boltzmann and generalized matching CFs (Fig. 10).

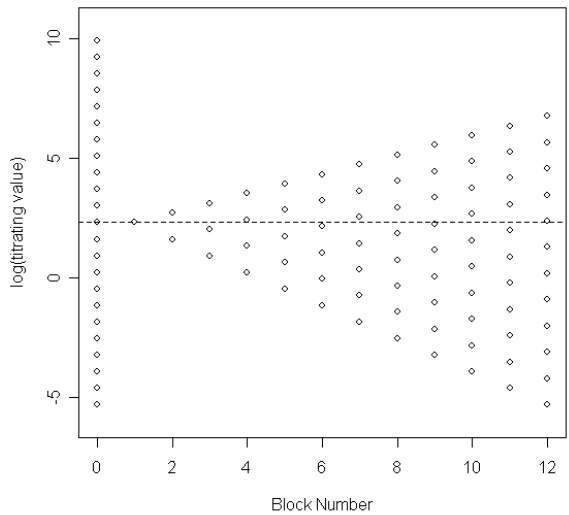

Fig. 12. State space under geometric adjustments. The y-axis shows the log of the titrating dimension and the x-axis shows the block number. Values for symmetric geometric adjustments are shown at $x = 0$ while $x > 0$ shows the evolution of the state space under asymmetric geometric adjustments. Values computed using an initial value of 10 and $\delta = 0.5$. The dashed line in the graph is $log(10)$, meant to represent the indifference point.

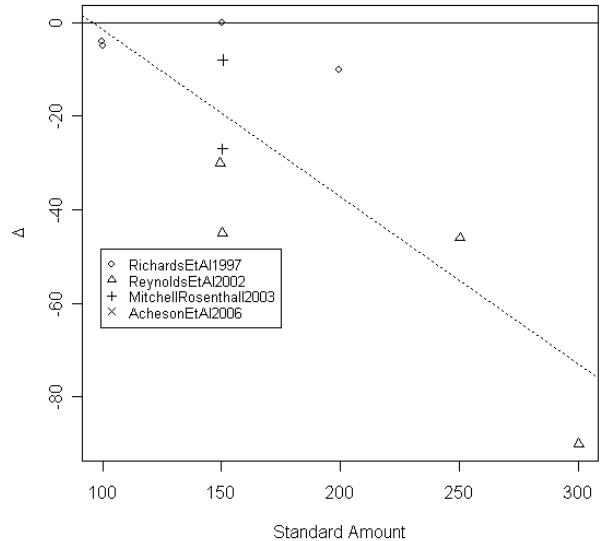

**Fig. 13.** Observed bias in experiments with asymmetrically adjusting amounts. All experiments used rats, and rewards are measured in µL of water. All experiments used the procedure from Richards et al. (1997). Solid line depicts no bias, and the dotted line shows the best fitting linear regression line. Indifference points were estimated from graphs if not reported numerically.

Although in their initial paper Richards et al. (1997) observed a slightly negative titration bias, they concluded that "the adjusting-amount procedure, used in conjunction with daily changes in the standard, can be used to rapidly determine discount functions for delay". However, this negative titration bias has been consistently replicated in following studies, often with much larger magnitudes.

## Discussion

Intuitively, the titration procedure seems like a reasonable method for obtaining utility equivalence, and indeed it will produce unbiased results if subjects decide deterministically. Despite this intuition, our analysis shows that the procedure will generate biased results for many simple learning models if subjects make decisions probabilistically, as non-human subjects do in titration experiments. This bias emerges even for more realistic learning models that base decisions on a EWMA of the rewards received, and there is little reason to expect the bias to disappear for other models of decision making. Thus, it is likely that the results obtained from titration experiments are biased and that inferences based on such results, including selection between different types of utility functions may be flawed.

Interestingly, this bias was identified in the first paper to popularize titration procedures (Mazur, 1984), but was treated as an intrinsic bias of the subjects against the titration procedure, rather than a direct result of the procedure itself. Because the bias observed in this first experiment increased approximately linearly with the value of the standard, Mazur (1984, 1988) proposed correcting for the bias in a general titration experiment by first conducting the experiment with identical rewards, estimating the ratio of $\bar{\tau}/s$, and then dividing the indifference points obtained under other conditions by this ratio. This correction method only works if the magnitude of the bias changes in the same manner in both experiments with identical and non-identical rewards. However, our simulation results show that the magnitude of the bias can change quite differently when the rewards are identical and when they are non-identical, a result that held across all adjusting procedures studied (compare Fig. 5 and Fig. 6 for example). If this result holds in general, the correction method proposed by Mazur (1984, 1988) will not eliminate the bias, and a more complex correction method, or a different titration procedure, will be needed.

Although the titration bias complicates the computation of utility equivalence point, the nature and magnitude of the bias also has the potential to help distinguish between different learning models. For example, the titration bias observed in Mazur (1984) is consistent with the Boltzmann CF, but not with the generalized matching CF. Conversely, the bias observed by Bateson & Kacelnik (1995) is not consistent with either of the CFs tested, suggesting a different UF is needed. Finally, the titration bias observed across experiments using asymmetric geometrically adjusting amounts is qualitatively consistent with the bias predicted under both the Boltzmann and generalized matching CFs (compare the top row of Fig. 11 and Fig. 13), but a closer quantitative comparison could potentially distinguish between them.

Nonetheless, the experimental results are somewhat equivocal on the whole, especially the results from experiments using arithmetically adjusting delays. The models presented here are incapable of generating the negative titration bias observed in Mazur (1986a) and Grace (1996), but are consistent with the other experiments that showed a positive titration bias (Fig. 7). The incongruity between these experiments deserves closer study.

Indeed, much would be gained from a study that examined the titration bias systematically. By presenting the subject with a series of conditions involving identical rewards, the simple models proposed above could be distinguished and possibly rejected. Moreover, the stability of the indifference points obtained with the titration procedure should be evaluated. Several experiments (Mazur, 1988; Wade et al., 2000) have analyzed how the value of the obtained indifference points depends on aspects of the titration procedure such as step size or starting value, but to our knowledge no experiments have examined whether the reward values at the indifference point actually lead to indifference in a non-titrating context. It may be that animals respond differently in changing environments, such as provided by the titration procedure, than they do in unchanging environments where reward contingencies are stable (Steele-Feldman, 2006). To test this possibility, subjects could be retested with the reward values at the indifference point after a delay of several weeks or with different operants representing each reward. If the subjects continue to show indifference, than we may conclude that the indifference points do truly represent utility equivalence points.

In any case, our analysis calls into question inferences about utility functions based on indifference points obtained with titration procedures. The titration procedure has also been questioned on substantially different grounds (Cardinal et al., 2002). Taken together, these two studies strongly suggest caution in interpreting the results of titration experiments.

# Appendix

## Arithmetic Adjustments

To establish conditions under which the expected titration bias $E(\Delta) = E(\bar{\tau}) - \tilde{\tau}$ is greater than zero first define

$$\beta_i = Y_i \Big/ \sum_{j=0}^{\infty} Y_j \qquad (16)$$

Using Eq. (13), we can then express the expected titration bias as

$$E(\Delta) = \sum_{i=0}^{\infty} (\tau_i - \tilde{\tau})\beta_i \qquad (17)$$

where $\tau_i = \delta i$. For reasons that will be clear later, we will re-index so that this equation with $\tau_i = \tilde{\tau} + \delta i$ and then

$$E(\Delta) = \sum_{i=-\tilde{\tau}/\delta}^{\infty} (\tau_i - \tilde{\tau})\beta_i . \qquad (18)$$

We can rewrite Eq.( 18) as

$$E(\Delta) = \sum_{i=1}^{\tilde{\tau}/\delta} \delta i (\beta_i - \beta_{-i}) + \sum_{i=\tilde{\tau}/\delta+1}^{\infty} (\tau_i - \tilde{\tau})\beta_i \qquad (19)$$

The second summation, which is clearly greater than or equal to zero, generates the residual titration bias, and the first summation, when it is strictly greater than zero, generates the robust positive titration bias.

To show that a given learning model will generate a robust positive titration bias, we must show that the first summation is strictly greater than zero. So if we can show that $\beta_i - \beta_{-i} > 0$ for $i = \{1, 2, ..., \tilde{\tau}/\delta\}$, we are done.

<u>Proposition 1:</u> If $\theta_0^+ = \theta_0^-$, $\theta_i^+ - \theta_{-i}^- > 0$, and $\theta_{-i}^+ - \theta_i^- > 0$ for $i = \{1, 2, ..., \tilde{\tau}/\delta\}$ then a learning model will display a robust titration bias.

<u>Proof:</u> The proof will proceed by induction. First note, from Eqs. (13) and (16), that

$$\beta_1 = \beta_0 \frac{\theta_0^+}{\theta_1^-} \text{ and } \beta_{-1} = \beta_0 \frac{\theta_0^-}{\theta_{-1}^+} . \qquad (20)$$

Subtracting these gives

$$\beta_1 - \beta_{-1} = \beta_0 \left( \frac{\theta_0^+}{\theta_1^-} - \frac{\theta_0^-}{\theta_{-1}^+} \right) = \beta_0 \frac{\theta_0^-}{\theta_1^-} \left( \frac{\theta_0^+}{\theta_0^-} - \frac{\theta_1^-}{\theta_{-1}^+} \right). \qquad (21)$$

If $\theta_0^+ = \theta_0^-$, this becomes

$$\beta_1 - \beta_{-1} = \beta_0 \frac{\theta_0^-}{\theta_1^-} \left( 1 - \frac{\theta_1^-}{\theta_{-1}^+} \right). \qquad (22)$$

Thus if $\theta_{-i}^+ > \theta_i^-$ for $i = 1$, then $\beta_1 - \beta_{-1} > 0$.

As above,

$$\beta_{i+1} - \beta_{-i-1} = \beta_i \frac{\theta_i^+}{\theta_{i+1}^-} - \beta_{-i} \frac{\theta_{-i}^-}{\theta_{-i-1}^+} \qquad (23)$$

Assuming that $\beta_i - \beta_{-i} > 0$, we obtain

$$\beta_{i+1} - \beta_{-i-1} > \beta_i \left( \frac{\theta_i^+}{\theta_{i+1}^-} - \frac{\theta_{-i}^-}{\theta_{-i-1}^+} \right) \qquad (24)$$

or equivalently

$$\beta_{i+1} - \beta_{-i-1} > \beta_i \frac{\theta_{-i}^-}{\theta_{i+1}^-} \left( \frac{\theta_i^+}{\theta_{-i}^-} - \frac{\theta_{i+1}^-}{\theta_{-i-1}^+} \right). \qquad (25)$$

So if $\theta_{-i}^- < \theta_i^+$ and $\theta_{-i}^+ > \theta_i^-$ for $i = \{1, 2, ... \tilde{\tau}/\delta\}$, then

$$\beta_{i+1} - \beta_{-i-1} > 0 . \qquad (26)$$

Thus by induction, $\beta_i - \beta_{-i} > 0$, and $E(\Delta) > 0$.

Now use this proposition to prove the theorems about difference-based choice functions, and then extend the theorems to apply to ratio-based choice functions using the difference based analogue:

<u>Theorem A.1:</u> With an attractive titrating dimension, models with a proper difference-based CF and a strictly concave UF will generate a robust titration bias.

<u>Proof:</u> For a balanced CF (Eq. (3)) if we know that $\theta_0^+ = \theta_0^-$ and that if $\theta_i^+ - \theta_{-i}^- > 0$ then necessarily $\theta_{-i}^+ - \theta_i^- > 0$. So the proof is complete if we can show that $\theta_i^+ - \theta_{-i}^- > 0$. From the definition of a strictly concave function

$$U(\tilde{\tau}) > \frac{U(\tilde{\tau} - i\delta) + U(\tilde{\tau} + i\delta)}{2} . \qquad (27)$$

Rearranging this equation gives

$$U(\tilde{\tau}) - U(\tilde{\tau} + i\delta) > U(\tilde{\tau} - i\delta) - U(\tilde{\tau}). \qquad (28)$$

Since the CF is proper and difference-based, it follows that

$$C(U(\tilde{\tau}), U(\tilde{\tau} + i\delta)) > C(U(\tilde{\tau} - i\delta), U(\tilde{\tau})), \qquad (29)$$

Recalling that $\tau_i = \tilde{\tau} + \delta i$, and using Eq. (10) gives

$$\sqrt{\theta_i^+} > \sqrt{\theta_{-i}^-} \qquad (30)$$

Squaring and rearranging, we obtain $\theta_i^+ - \theta_{-i}^- > 0$.

<u>Theorem A.2:</u> With an unattractive titrating dimension, models with a proper difference-based CF and a strictly convex UF will generate a robust titration bias.

<u>Proof:</u> The proof proceeds almost exactly as the previous one. From the definition of a strictly convex function

$$U(\tilde{\tau} + i\delta) - U(\tilde{\tau}) > U(\tilde{\tau}) - U(\tilde{\tau} - i\delta). \qquad (31)$$

As in the previous proof it then follows from the definition of a proper difference-based CF and from Eq. (11) that $\theta_i^+ - \theta_{-i}^- > 0$.

<u>Theorem A.3:</u> With either an attractive or unattractive titrating dimension, a proper difference-based CF and an affine UF will not generate a robust titration bias.

<u>Proof:</u> From Eq. (19), the proof is complete if we can show that $\beta_i = \beta_{-i}$ for $i = \{1, 2, ..., \tilde{\tau}/\delta\}$. By an argument similar to that presented in Proposition 1, this reduces to showing that $\theta_0^+ = \theta_0^-$ and $\theta_{-i}^+ = \theta_{+i}^-$. Again by the definition of a balanced CF, we know that $\theta_0^+ = \theta_0^-$. For an affine function,

$$U(\tilde{\tau}) - U(\tilde{\tau} + i\delta) = U(\tilde{\tau} - i\delta) - U(\tilde{\tau}). \qquad (32)$$

For a proper difference-based CF and an attractive titrating dimension this implies that $\theta_{-i}^+ = \theta_i^-$ by following the argument in Theorem A.1. Similarly, for an unattractive titrating dimension we can rearrange Eq. (32) to show that $\theta_{-i}^+ = \theta_i^-$ by following the argument in Theorem A.2.

*Ratio-based choice functions*

The preceding proofs can be applied to models with ratio based CFs by defining $U_L(i) = \log(U(i))$, and treating $U_L(i)$ as the UF for the difference based analogue to the ratio based CF, $U(i)$. If $U(i)$ is strictly log-concave, it follows that $U_L(i)$ is strictly concave and thus the proof in Theorem A.1 applies. Similarly Theorem A.2 can be used if $U(i)$ is strictly log-convex. Finally, if $U(i) = e^{a+bi}$ where $a$ and $b$ are real constants, (i.e. $U(i)$ is log-affine) then $U_L(i)$ is an affine function, and the proof in Theorem A.3 holds.

## Symmetric Geometric Adjustments

With symmetric geometric adjustments, we must make two modifications to the formalism introduced for arithmetic adjustments. First of all, because the adjusting rule always involves multiplication or division by a fraction, the geometric procedure has no firm lower or upper bound, and although the value of the titrating dimension remains confined to $(0, \infty)$, the states range as $i \in (-\infty, +\infty)$, and the values accessible in the geometric adjustment procedure depend on the initial value. For simplicity, we will assume that the initial value used at the beginning of the procedure is equal to the utility equivalence point $\tilde{\tau}$ in which case we can index the states as $\tau_i = \tilde{\tau}\delta^{-i}$ with $\delta \in (0,1)$. Second, we must modify Eq. (13) to take into account the new state space[3]:

$$E(\bar{\tau}) = \lim_{N \to \infty} E(\tau(N)) = \sum_{i=-\infty}^{+\infty} \tau_i Y_i \Big/ \sum_{i=-\infty}^{+\infty} Y_i$$
$$Y_i = \prod_{j=-\infty}^{i-1} \frac{\theta_j^+}{\theta_{j+1}^-} \qquad (33)$$

Defining $\beta_i$ as in the previous section, we can then write the titration bias as

$$E(\Delta) = \sum_{i=-\infty}^{+\infty} (\tau_i - \tilde{\tau})\beta_i = \sum_{i=-\infty}^{+\infty} (\tilde{\tau}\delta^{-i} - \tilde{\tau})\beta_i \qquad (34)$$

---

[3] To compute this sum, we must choose some smallest state where the probability of reaching the state from the starting value is ~ zero, and then set in this state. This minor point is irrelevant to the following analysis.

A bit of arithmetic then gives

$$E(\Delta) = \sum_{i=1}^{\infty} \tilde{\tau}\left(\delta^{-i}-1\right)\left(\beta_i - \delta^i \beta_{-i}\right) \quad (35)$$

Since the term in the first parenthesis is positive and $\delta^i < 1$ for $i > 1$, a learning model generates a robust titration bias if $\beta_i - \beta_{-i} \geq 0$ for $i = \{1, 2, ..., \infty\}$. Proposition 1 is then applicable, and we need only derive appropriate conditions to re-prove the associated theorems with arithmetic adjustments. All arithmetic conditions are based on the definition of a convex function. The proofs in this section are based on a generalized notion of convexity, Geometric-Arithmetic (GA) convexity (Niculescu, 2003). For any GA-convex function $U$ it holds that

$$U(\tilde{\tau}) \leq \frac{U(\tilde{\tau}\delta^i) + U(\tilde{\tau}\delta^{-i})}{2}. \quad (36)$$

Condition of Eq. (36) can be used to prove the following theorems in a manner directly analogous to the corresponding theorems for arithmetic adjustments.

<u>Theorem G.1:</u> With an attractive titrating dimension, models with a proper difference-based CF and a strictly GA-concave UF will generate a robust titration bias.

<u>Theorem G.2:</u> With an attractive titrating dimension, models with a proper difference-based CF and a strictly GA-convex UF will generate a robust titration bias.

# Acknowledgement

This work was supported in part by the Bonneville Power Administration and the ARCS foundation Lee and Mike Brown fellowship. We thank Jerker Denrell, Peter Killeen, and several anonymous reviewers for helpful comments.20